\documentstyle[12pt,psfig]{article}
\begin{document}

\begin{center}
\large{{\bf Search for periodic vacuum in QED$_2$}}
\end{center}
\begin{center}
{S. Nagy and K. Sailer}
\end{center}
\begin{center}
{{\em Department for Thoretical Physics, 
Kossuth Lajos University,
Debrecen, Hungary}}
\end{center}

\begin{abstract}
It is shown that the vacuum of QED$_2$  in Minkowski spacetime 
does not favour a periodic electric mean field.
The projected  effective action exhibiting a genuine dependence on the
non-vanishing background field has been introduced.
The functional dependence of the energy density of the vacuum on 
the assumed  periodic  vacuum expectation value of the vector potential
is determined from the component $T^{00}$ of the energy-momentum 
tensor at one-loop order. 
Treating the background field non-perturbatively,
the energy of the vacuum in the presence of a periodic mean field is
found not be equal to the negative of the effective action.
\end{abstract}

\section{Introduction}\label{Introduction}

It is well-known that the perturbative vacuum of QED is trivial,
i.e. it is characterized by the vanishing expectation value of the
electromagnetic vector potential. Our goal is to investigate
the vacuum non-perturbatively in the framework of the continuum 
theory in Minkowski spacetime. As a working hypothesis we assume the
presence of a periodic electric background  field. The interaction of
the fermions with the background field is treated exactly by
integrating out the fermion fields, whereas the quantum fluctuations
of the electromagnetic field are taken into account at one-loop order.
The non-perturbative treatment of the interaction with the background
introduces infinitely many non-renormalizable, i.e. irrelevant
interaction vertices. It may happen that such vertices generate
non-trivial, periodic (otherwise anti-ferromagnetic) vacuum structure.
Similar examples are known for Yang-Mills theories on the lattice
\cite{Fin97}, where the existence of various anti-ferromagnetic vacua  
have been established due to irrelevant interaction terms.

\begin{figure}[ht]
\hspace{1cm}
\psfig{file=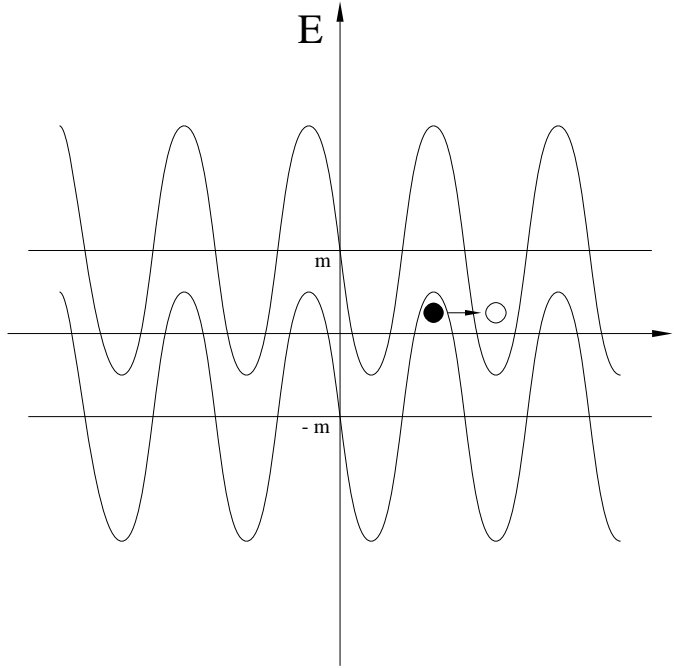,width=10cm}
\vspace{-1cm}
\caption[]{Intuitive picture of pair creation in overcritical periodic 
external electric field}
\label{tunnel}
\end{figure}

Our problem setting can also be motivated by the following intuitive
picture. In the presence of a background scalar potential 
${\bar A}_0
= a \cos (\ell x)$ with overcritical amplitude $a >m$ (with the rest
 mass of the electron $m$) localized electrons of the Dirac sea may
 tunnel into localized electron states of the same energy above the
mass gap (see Fig. \ref{tunnel}). Therefore, the creation of a certain
 amount of electron-positron
pairs with the spatial separation of $\pi/\ell$ is imaginable. This 
might stabilize itself with a periodic charge density (but zero net
charge) and a periodic mean electric field. 
The expected effect might be destroyed due to the uncertainty
principle. Localisation in an interval $\Pi/2\ell$ means a momentum
and, consequently, an energy spread of order 
$\Delta E=\frac{p}{E}\Delta p $ and the effect may occur only if 
$a>>m+\frac{\Delta E}2$.

\section{Projection method\label{projection}}

It is not completely trivial to define an effective action depending
on a background field ${\bar A}^\mu (x)$. The simple shift of the
integration variable $A^\mu (x)$ to $\alpha^\mu (x) = A^\mu (x)$ -
${\bar A}^\mu (x)$ is not the answer requested, since a dependence on
the background ${\bar A}^\mu (x)$ shall only occur due to terminating
the loop expansion. Below we introduce an effective action exhibiting
a genuine dependence on the background field.

Our method of defining the sector of QED for field configurations
belonging to quantum fluctuations around a given vacuum expectation
value $\langle A^\mu (x) \rangle$ consists of the following steps:
\begin{enumerate}
\item {\em Projection to the sector with a given background field.}
The generating functional of QED in Minkowski spacetime has the form:
\begin{eqnarray}
   Z_{QED} &=&
    \int {\cal D} {\bar \psi} {\cal D} \psi {\cal D} A 
   \exp \left\{ {\rm i} S_{em} \lbrack A, \xi \rbrack
      +  {\rm i} S_D \lbrack A, {\bar \psi}, \psi \rbrack 
         \right\}
\end{eqnarray}

with the vector potential $A^\mu (x)$ $(\mu = 0,1,2,3)$ and the
electron-positron field $\psi (x)$, where $S_{em} \lbrack A, \xi \rbrack $
and $ S_D \lbrack A, {\bar \psi}, \psi \rbrack$ are the action of the
electromagnetic field and the Dirac action, resp., in the covariant
gauge with the gauge parameter $\xi$. Let the vector $n^\mu (x)$
be introduced in the space of the vector potential configurations.
Multiply the integrand of the generating functional by the factor 1
written in the form
\begin{eqnarray}
 1 &=& \int dc \ \delta ( \int dx A_\mu n^\mu - c \Omega)
\end{eqnarray}
with $\Omega = TV$ the spacetime volume. Then, we find
\begin{eqnarray}
  Z_{QED} &=& \int dc Z_{QED} ' \lbrack n,c \rbrack ,
\end{eqnarray}
where the functional
\begin{eqnarray}
   Z_{QED} ' \lbrack n,c \rbrack &=&
    \int {\cal D} {\bar \psi} {\cal D} \psi {\cal D} A 
   \exp \left\{ {\rm i} S_{em} \lbrack A, \xi \rbrack
      +  {\rm i} S_D \lbrack A, {\bar \psi}, \psi \rbrack 
        \right\}
               \nonumber\\
        &   &
    \delta  ( \int dx A_\mu n^\mu - c \Omega)
\end{eqnarray}
can be concieved as the generating functional of the sector
belonging to vector potential configurations in a hypersurface
orthogonal to $n^\mu (x)$. The projected generating functional
$  Z_{QED} ' \lbrack n,c \rbrack$ is gauge invariant for the choice of
$n^\mu (x)$ satisfying the condition $\partial_\mu n^\mu (x) =0$.
The projected effective action $S_{eff} \lbrack n,c \rbrack$  is defined by
\begin{eqnarray}
\label{seffdf}
    Z_{QED} ' \lbrack n,c \rbrack &=& 
       \exp \left\{ {\rm i} S_{eff} \lbrack n,c \rbrack
            \right\} .
\end{eqnarray}

For the choice $n^\mu (x) = {\bar A}^\mu (x)$ with a given
background field ${\bar A}^\mu (x)$ satisfying the Lorentz condition
$\partial_\mu {\bar A}^\mu (x) =0$, we obtain the effective action for
the sector belonging to the given background  ${\bar A}^\mu (x)$.
Introducing the shifted integration variable $\alpha^\mu (x) =
A^\mu (x) - {\bar A}^\mu (x) $, the projected generating functional 
can be rewritten as
\begin{eqnarray}
\label{zqedp}
    Z_{QED} ' \lbrack {\bar A},c \rbrack &=&
  \int d\lambda \int {\cal D} \alpha 
  \exp \left\{ {\rm i} S_{em} \lbrack {\bar A} + \alpha , \xi \rbrack
       \right\}
       \nonumber\\
     &   &
    \exp \left\{  
   {\rm i} \lambda \left( \int dx {\bar A}^\mu {\bar A}_\mu + \int dx {\bar
     A}^\mu \alpha_\mu - c \Omega \right)
           \right\}
      \nonumber\\
  &   &
   \int {\cal D} {\bar \psi} {\cal D} \psi
    \exp \left\{ {\rm i} S_D \lbrack {\bar A} + \alpha , {\bar \psi}
      , \psi \rbrack  \right\} .
\end{eqnarray}

\item {\em Identification of the background with the vacuum
    expectation value of the vector potential.}
As to the next, it is required that
\begin{eqnarray}
\label{constraint}
  \langle A^\mu (x) \rangle = {\bar A}^\mu (x)
        , \qquad {\mbox {i.e.}} \qquad
    \langle \alpha^\mu \rangle =0 .
\end{eqnarray}
This condition is used to determine the constant $c$ as the functional
of the vacuum expectation value of the vector potential
 $  \langle A^\mu (x) \rangle$.
\end{enumerate}

For later use it is useful to introduce the external sources $j^\mu
(x)$, ${\bar \eta} (x)$, and $\eta (x)$ coupled to the quantum
 fluctuation $\alpha^\mu (x)$ of the vector
potential, and to the fermion fields $\psi (x)$ and ${\bar \psi} (x)$,
resp. Then, we find instead of Eq. (\ref{zqedp}) the expression
\begin{eqnarray}
\label{zqedpj}
    Z_{QED} ' \lbrack {\bar A},c , j, \eta , {\bar \eta} \rbrack &=&
  \int d\lambda \int {\cal D} \alpha 
  \exp \left\{ {\rm i} S_\lambda \lbrack {\bar A} + \alpha,
    \xi, c    \rbrack
     + \int dx j^\mu \alpha_\mu                               
          \right\}
      \nonumber\\
   &   &
     Z_F \lbrack A, \eta, {\bar \eta} \rbrack
\end{eqnarray}
with
\begin{eqnarray}
  S_\lambda \lbrack {\bar A} + \alpha, \lambda
    ,\xi, c    \rbrack
          &=&
 S_{em} \lbrack {\bar A} + \alpha , \xi \rbrack
+   \lambda \left( \int dx {\bar A}^\mu {\bar A}_\mu + \int dx {\bar
     A}^\mu \alpha_\mu - c \Omega \right) ,
\end{eqnarray}
and
\begin{eqnarray}
    Z_F \lbrack A, \eta, {\bar \eta} \rbrack
          &=&
     \int {\cal D} {\bar \psi} {\cal D} \psi
    \exp \left\{ {\rm i} S_D \lbrack {\bar A} + \alpha , {\bar \psi}
      , \psi \rbrack  \right\} .
\end{eqnarray}

\section{Projected effective action}

The projected generating functional (\ref{zqedp}) defines the
projected effective action via Eq. (\ref{seffdf}). We determine it in
one-loop approximation, i.e. we replace the full vector potential
$A^\mu (x)$ in the Dirac action $S_D$ by the background field ${\bar
  A}^\mu (x)$. Thus, the generating functional is factored into an
electromagnetic part and a fermionic part,
\begin{eqnarray}
     Z_{QED} ' \lbrack {\bar A},c , j, \eta , {\bar \eta} \rbrack
 &=&
  Z_{em} ' \lbrack {\bar A},c , j \rbrack 
   Z_F \lbrack {\bar A}, \eta, {\bar \eta} \rbrack .
\end{eqnarray}
The explicit forms of the actions are given as
\begin{eqnarray}
   S_{em } \lbrack {\bar A}+\alpha , \xi \rbrack
  &=&
  \frac{1}{2} ( {\bar A} D^{-1} {\bar A} ) 
    +  \frac{1}{2} ( \alpha D^{-1} \alpha ) 
    +  ( \alpha D^{-1}  {\bar A} )  ,
\end{eqnarray}
and
\begin{eqnarray}
   S_D \lbrack {\bar A} \rbrack &=&
     ( {\bar \psi} G^{-1} \psi ) 
\end{eqnarray}
with the inverse of the photon propagator in Lorentz gauge,
\begin{eqnarray}
   D^{-1}_{\mu \nu} (x,y) &=& 
    \left\lbrack  g_{\mu \nu} {\Box}_x +
          \left( \xi^{-1} - 1 \right) \partial^x_\mu \partial^x_\nu
     \right\rbrack \delta (x-y) ,
\end{eqnarray}
and the inverse of the fermion propagator in the background field
${\bar A}^\mu (x)$,
\begin{eqnarray}
  G^{-1} (x,y) &=&  \left( {\rm i} \gamma^\mu (\partial^x_\mu 
         - i \bar A_\mu (x) )  - m \right) \delta (x-y) 
\end{eqnarray}
with the Dirac matrices $\gamma^\mu$   $( \mu = 0,1,2,3)$.
For the sake of simplicity, the notation $( f  O g)= \int dx dy
f^a(x) O_{ab }(x,y) g^b (y)$ is used, where $a$ and $b$ are either
Lorentz
or spinor summation indices. In the one-loop approximation the path
integrals are Gaussian ones and can be performed explicitly,
leading to
\begin{eqnarray}
 \ln Z_{em} ' \lbrack {\bar A},c , j \rbrack 
          &=&
   - \frac{1}{2} {\mbox { Tr }} \ln D^{-1} 
      - \frac{1}{2} \ln ( {\bar A} D {\bar A} ) 
     + \frac{ \rm i}{2} (jD_1 j)   - (j {\bar A} )
               \nonumber\\\
      &   &  
     + \frac{i}{2}  \frac{ c^2 \Omega^2}{ ( {\bar A} D {\bar A} ) }
    + c \Omega \frac{ (j D {\bar A} ) }{ ( {\bar A} D {\bar A} )  } ,
\end{eqnarray}
and
\begin{eqnarray}
    \ln Z_F \lbrack {\bar A}, \eta, {\bar \eta} \rbrack 
     &=&
    {\mbox { Tr }} \ln G^{-1} + {\rm i} ( {\bar \eta} G \eta )
\end{eqnarray}
with the modified photon propagator
\begin{eqnarray}
\label{modpro}
   D_1^{\mu \nu} (x,y) 
            &=&
    D^{\mu \nu} (x,y) - \frac{ \int du D^{\mu \rho}(y,u) {\bar A}_\rho
      (u)  \int dv D^{\nu \sigma}(x,v) {\bar A}_\rho (v)}{
          ( {\bar A} D {\bar A} )   } .
\end{eqnarray}

Now we restrict our considerations to $1+1$ dimensional systems and
time independent periodic backgrounds of the form
\begin{eqnarray}
\label{ansatz}
{\bar A}^\mu = \delta ^{0\mu} a \cos (\ell x_1 )
\end{eqnarray}
 satisfying the
Lorentz condition.

The constant $c$ can be determined by using the fact that
\begin{eqnarray}
\label{cnull}
0 = \int {\cal D} \alpha \int d\lambda
   \frac{\partial}{\partial\lambda}
     \exp \left\{ {\rm i} S_\lambda \lbrack {\bar A} + \alpha,
    \xi, c    \rbrack
     + \int dx j^\mu \alpha_\mu                               
          \right\}.
\end{eqnarray}
From this we find that
\begin{equation}
c_0\Omega=\int dx \bar A_\mu \bar A^\mu.
\end{equation}

One establishes now for the one-loop effective action:
\begin{eqnarray}
\label{effac}
  {\rm i} S_{eff}& =& 
          \ln  Z_{em} ' \lbrack {\bar A},c_0 , j=0 \rbrack  
   +    \ln Z_F \lbrack {\bar A}, \eta=0, {\bar \eta}=0 \rbrack 
                 \nonumber\\
    & = &  
      {\mbox { Tr }} \ln G^{-1} - \frac{1}{2}{\mbox { Tr }} \ln D^{-1}
    + \Omega \frac{\rm i}{4} a^2 \ell^2 - \frac{1}{2} \ln
    \frac{a^2}{\ell^2} .
\end{eqnarray}

In the infinite volume limit one finds
\begin{eqnarray}
\label{actden}
 - \Omega^{-1} S_{eff} & \sim &
   - V^{-1} {\sum_{k}}^>  \epsilon_k  + V^{-1} \sum_k \frac{1}{2} \omega_k
        - \frac{1}{4} a^2 \ell^2 .
        \end{eqnarray}

\begin{figure}[ht]
\hspace{1cm}
\psfig{file=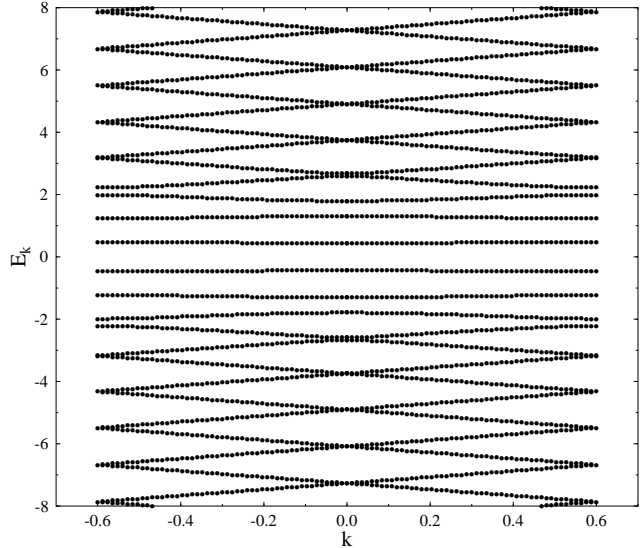,width=10cm}
\vspace{-1cm}
\caption[]{Energy eigenvalues with changing momentum}
\label{felhasad}
\end{figure}

Here the first and the second terms represent the energy density of
the Dirac vacuum and that of the free electromagnetic field, resp. 
(see e.g. \cite{Raj89}). $\sum^>$ denotes the summation over all
non-negative single fermion energies $\epsilon_k \ge 0$, being the energy
eigenvalues of the Dirac equation in the external field ${\bar A}^\mu (x)$.
 The third term on the r.h.s. of Eq. (\ref{actden})
 is just the negative of the 
energy density of the periodic electric background field. The last term of the
effective action (\ref{effac}) gives a vanishing contribution to the 
action density in the infinite volume limit.

 The meaning of the first two terms of Eq. (\ref{actden})
 might lead one to the false conclusion that the negative of the action density is
 equal to the energy density of the vacuum, as it would happen if the
background field were constant \cite{Raj89}. Determining
the energy density from the energy-momentum tensor we will show
that the energy
density does not equal the negative of the action density for
inhomogeneous background field. 

\begin{figure}[htb]
\hspace{1cm}
\psfig{file=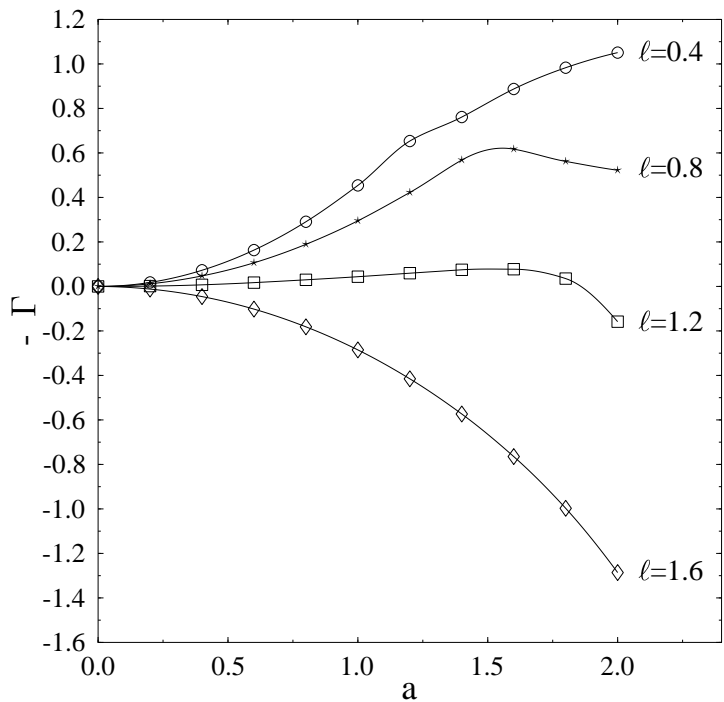,width=10cm}
\vspace{-1cm}
\caption[]{The difference of the negative of the effective actions in the
  presence and in the absence of a periodic background field for
  various values of $a$ and $\ell$.
}
\label{effacden}
\end{figure}

As 1+1 dimensional QED is a superrenormalizable theory, the action
density is UV finite after subtracting the action density of the free
vacuum $S_{eff,free}$, i.e. that of the vacuum in the absence of 
the background field.
This difference $-\Gamma=\Omega^{-1}(-S_{eff}+S_{eff,free})$
has been calculated numerically by solving the Dirac
equation for the single-fermion energies. 

These energy eigenvalues show the same band structure plotted against the 
momentum in fig. \ref{felhasad}
as the electrons in the Kronig-Penney model \cite{Lan81} because
of the periodic potential.
Fig. \ref{effacden} shows how $-\Gamma$ changes by modifying the
amplitude $a$ and the wavelength $\ell$ of the potential. 
(The fermion rest mass is chosen for $m{=}1$.)
One can recapitulate from fig. \ref{effacden} that the surface 
$-\Gamma(a,\ell)$ has only a single stationary point at the
origin of the parameter space $(a, \ell)$, i.e. the path integral
defining the vacuum-vacuum transition amplitude is dominated by
the trivial, identically vanishing field configuration $A^\mu (x) =0$.

\section{Mean field equation}
The vacuum expectation value of the current can be written at one-loop
order in the following form:
\begin{eqnarray}
\label{aram}
  \langle {\bar \psi} (x) \gamma^\mu \psi (x) \rangle
   &=&
   \int d\lambda \int {\cal D}\alpha 
     e^{ {\rm i} S_\lambda }
 \frac{ \delta }{ {\rm i} \delta {\bar A}_\mu (x) } Z_F  
          \left/ 
    \int d\lambda \int {\cal D}\alpha 
     e^{ {\rm i} S_\lambda }  Z_F  \right.
                 \nonumber\\
   &  = &
  e^{ - {\rm i} S_{eff} } \left\lbrack 
    \frac{ \delta }{ {\rm i} \delta {\bar A}_\mu (x) }
            e^{  {\rm i} S_{eff} }
       - Z_F    \frac{ \delta }{ {\rm i} \delta {\bar A}^\mu (x) }
           Z_{em} '  \right\rbrack .
\end{eqnarray}
The first term on the r.h.s. of Eq. (\ref{aram}) is just the first
functional derivative of the effective action $\delta S_{eff}/\delta
{\bar A}_\mu (x) $. In the infinite volume limit the second term gives:
\begin{eqnarray}
  &-& \exp \left\{ - \frac{ {\rm i} c_0^2}{ 2 ( {\bar A} D {\bar A} ) }
          \right\}
   \frac{ \delta }{ {\rm i} \delta {\bar A}_\mu (x) }   
\exp \left\{ \frac{ {\rm i} c_0^2}{ 2 ( {\bar A} D {\bar A} ) }
          \right\}
   \nonumber\\
 &=& 
  - \frac{ {\bar A}^\nu (x) }{ (D{\bar A} )^\nu (x)   }
  \left( \frac{ \delta c_0 }{ \delta {\bar A}_\mu (x) } 
          - {\bar A}^\mu (x) \right)
          \nonumber\\
  &=& - \ell^2 {\bar A}^\mu (x) = \partial_\nu {\bar F}^{\mu \nu}
\end{eqnarray}
with the field strength tensor $ {\bar F}^{\mu \nu}$ evaluated from
the background ${\bar A}^\mu$.
Here we used that 
\begin{eqnarray}
  \frac{ \delta c_0 }{ \delta {\bar A}_\mu (x) }
           &=&
  2 {\bar A}^\mu (x) 
\end{eqnarray}
due to Eq. (\ref{cnull}). Thus, one obtains the equation
\begin{eqnarray}
\label{mfeqact}
  \langle  {\bar \psi} (x) \gamma^\mu \psi (x) \rangle
     &=& 
     \frac{ \delta S_{eff} }{  \delta {\bar A}_\mu (x) }
   +  \partial_\nu {\bar F}^{\mu \nu}.
\end{eqnarray}
For the background field configuration making the effective action
extremum, one recovers the vacuum expectation value of the Maxwell
equation. Thus,  the effective action has an extremum at the
background field configuration coinciding with the mean field solution.
In our numerical search, this is the trivial extremum found at
$a=\ell=0$.

Satisfying the necessary condition of the extremum of the effective
action, Eq. (\ref{mfeqact}) results in the Poisson's equation for the
mean field $\bar A^0$, which must be considered together with coupled
set of operator equations for the quantum fields $\psi(x)$, $\bar\psi(x)$
and $\alpha^\mu(x)$. The latter equations should be solved and the
result substituted in the l.h.s. of Eq. (\ref{mfeqact}), in order to
make the charge density explicit.

\section{Energy density of the vacuum}

The symmetric energy momentum tensor is defined by \cite{Lan81}:
\begin{eqnarray}
  T^{\mu \nu} &=& 
   \frac{ \partial {\cal L} }{\partial F_{\kappa \mu} }
     F_\kappa^{\; \nu} 
  + \frac{ \partial {\cal L} }{\partial ( D_\mu \psi ) }
         D^\nu \psi  
   + D^{\nu \star}\psi \frac{ \partial {\cal L} }{ \partial
     (D_\mu^\star {\bar \psi} )                 }
   - g^{\mu \nu} {\cal L}
\end{eqnarray}
with the Lagrange density
\begin{eqnarray}
   {\cal L} &=& - \frac{1}{4} F_{\mu \nu} F^{\mu \nu} - \frac{1}{2\xi}
   (\partial_\mu A^\mu )^2 
    + \lambda A_\mu {\bar A}^\mu - \lambda c_0
     \nonumber\\
   &   &
   + \frac{i}{2} \left( {\bar \psi} \gamma^\mu D_\mu \psi 
        - D_\mu^\star {\bar \psi} \gamma^\mu \psi \right)  - m {\bar
        \psi} \psi
\end{eqnarray}
corresponding to the action $\int dx {\cal L} = S_\lambda + S_D$
and with the covariant derivative $D^\mu = \partial^\mu - {\rm i}
A^\mu $. 

Substituting the ansatz (\ref{ansatz}) one obtains for
the component $T^{00}$ of the energy-momentum tensor:
\begin{eqnarray}
  T^{00} &=& T^{00}_{em,2} \lbrack  \alpha \rbrack
     + T^{00}_{em,1} \lbrack {\bar A}, \alpha , \lambda \rbrack
     + T^{00}_{em,0} \lbrack {\bar A} \rbrack 
     +  T^{00}_\lambda  \lbrack {\bar A}, c \rbrack 
        \nonumber\\
    &   &
  +T^{00}_F \lbrack {\bar A}+\alpha , {\bar \psi}, \psi \rbrack
\end{eqnarray}
where $T^{00}_{em, a}$ $(a=0,1,2)$ are the terms independent of the
fermion field and being of the order $(\alpha )^a$. Due to the
constraint (\ref{constraint}), the expectation value of the
first order term vanishes, therefore it can be neglected.
The second order term
\begin{eqnarray}
T^{00}_{em,2}  \lbrack  \alpha \rbrack &=&
   - \partial_0 \alpha_\mu \partial^0 \alpha^\mu +
   2 \partial_0 \alpha_\mu \partial^\mu \alpha^0
   - \partial_\mu \alpha_0 \partial^\mu \alpha^0
           \nonumber\\
    &   &
  + \frac{1}{2} \partial_\mu \alpha_\nu \partial^\mu \alpha^\nu 
  - \frac{1}{2} \partial_\mu \alpha_\nu \partial^\nu \alpha^\mu
  + \frac{1}{2\xi} \partial_\mu \alpha^\mu \partial_\nu \alpha^\nu
\end{eqnarray}
is the expression for the free electromagnetic field,
whereas the zeroth order term is given by
\begin{eqnarray}
\label{fielden}
   T^{00}_{em,0} \lbrack {\bar A} \rbrack 
    &=&
        - \partial_\mu {\bar A}_0 \partial^\mu {\bar A}^0
       + \frac{1}{2} \partial_\mu {\bar A}_\nu 
           \partial^\mu {\bar A}^\nu 
     \nonumber\\
   &=&
  \frac{1}{2} \left( \nabla_j {\bar A}^0 \right)^{\; 2}  
   =
  \frac{1}{2} \ell^2 a^2 \sin^2 (\ell x_1) 
\end{eqnarray}
and represents the energy density of the periodic background field.
Due to the projection to a particular sector of the theory, the
additional term 
\begin{eqnarray}
    T^{00}_\lambda  [{\bar A}, c]
      &=& 
   ( c_0 - {\bar A}^\mu (x){\bar A_\mu }(x) ) \lambda \nonumber\\
    &=& 
   \biggl(( \Omega^{-1} \int du  {\bar A}^\mu (u){\bar A_\mu }(u) 
      - {\bar A}^\mu (x){\bar A_\mu }(x) \biggr) \lambda
     \nonumber\\
   &=&
   \frac{1}{2} a^2 ( 1 - 2 \cos^2 (\ell x_1 ) ) \lambda
\end{eqnarray}
occurs.
It is easy to see that this term does not contribute to the energy,
but gives a non-vanishing periodic contribution to the energy density.
Finally, the term 
\begin{eqnarray}
\label{kinen}
     T^{00}_F &=& - \frac{i}{2}\partial^0 \left( {\bar \psi} \gamma^0 \psi
     \right) 
        + {\bar \psi} \gamma^0 {\tilde H_D} ( {\bar A} + \alpha )
            \psi 
\end{eqnarray}
represents the contribution of the fermions to the energy density.
The first term on the r.h.s. gives vanishing contribution to the total energy
if charge conservation is required, whereas the second term accounts for the
kinetic energy of the fermion system.
At one-loop order, we have to substitute $\alpha^\mu =0$ in the
`Hamilton operator', i.e. write
\begin{eqnarray}
     {\tilde H}_D ( {\bar A} )
 &=& - {\rm i} \gamma^0 \gamma^j \partial^x_j 
       - \gamma^0 \gamma^j {\bar A}_j (x) 
       + \gamma^0 m  .
\end{eqnarray}  

Since ${\bar A}_j(x) = 0$ in our case, we obtain ${\tilde H}_D ({\bar A}) = 
- {\rm i} \gamma^0 \gamma^j \partial^x_j + \gamma^0 m$, i.e. the free 
`Hamilton operator'.

The vacuum expectation value of $T^{00}$ in the presence of the background 
field ${\bar A^{\mu}}$ is defined as

\begin{eqnarray}
\label{37}
   \langle T^{00} (x) \rangle 
       &=&
  \int d\lambda \int {\cal D} \alpha  
  {\cal D} {\bar \psi} {\cal D} \psi  T^{00} (x) 
      e^{ {\rm i} \left( S_\lambda + S_D \right) }
   \left/  
  \int d\lambda \int {\cal D} \alpha  
  {\cal D} {\bar \psi} {\cal D} \psi
 e^{ {\rm i} \left( S_\lambda + S_D \right) }
         \right.
     \nonumber\\
\end{eqnarray}
Similarly, the vacuum expectation value of $T^{00}_{free}$ in the absence of
the background field  $\langle T^{00}_{free} (x) \rangle_{0}$ is given by 
substituting ${\bar A}_{\mu} (x) =0$ in Eq.(\ref{37}) and replacing 
$T^{00}(x)$ by  $T^{00}_{free}(x)$. Then, the Casimir energy of the vacuum 
due to the background field ${\bar A}_{\mu} (x) \neq 0$ is 
\begin{eqnarray}
\label{39}
E_c &=& \int dx_1 \langle T^{00} (x_1) \rangle - 
      \int dx_1 \langle T^{00}_{free} (x_1) \rangle_{0}\nonumber\\ 
&=& \int dx_1 \left[(l^2 a^2/2) \sin^2(l x_1) + 
  \langle {\bar \Psi} \gamma^0 H_{D0} \Psi \rangle -  
  \langle {\bar \Psi} \gamma^0 H_{D0} \Psi \rangle_{0} \right]
\end{eqnarray}
This expression of the Casimir-energy reminds one on the expression for
the energy of a system of electric charges in classical electrodynamics.
There, the energy is the sum of the energy of the electromagnetic field and 
the kinetic energy of the charges \cite{J}. In our case the Casimir energy
is the sum of the background electric field and the change of the relativistic 
kinetic energy (including the rest mass) of the Dirac-sea due to the 
presence of the background field.

Eq.(\ref{37}) can \ be \ rewritten \ by \ the \ help \ of the 
generating functional
$Z'_{QED} \lbrack {\bar A}, c, j, \eta, {\bar \eta} \rbrack$ as
\begin{eqnarray}
    \langle T^{00} (x) \rangle 
       &=&
  \left( 1/  Z'_{QED} \lbrack {\bar A}, c_0 \rbrack \right)  
     \left.
 T^{00}_{op} (x)
    Z'_{QED} \lbrack {\bar A}, c, j, \eta, {\bar \eta} \rbrack 
     \right|_{ j, {\bar \eta}, \eta =0; c=c_0 }
\end{eqnarray}
where $ T^{00}_{op} (x)$ denotes the operator obtained from $T^{00}$
by replacing the fields $\alpha^\mu (x)$, $\psi (x)$, and
${\bar \psi} (x)$ by  the operators $\delta/\delta j_\mu (x)$, 
$\delta / \delta {\bar \eta} (x) $, and $- \delta /\delta \eta (x)$
and the variable $\lambda$ by $ {\rm i} \Omega^{-1} \partial /\partial c$.

Then, we find
\begin{eqnarray}
  \langle T^{00}_\lambda \rangle &=&
  \frac{1}{2} a^2 \ell^2 \left(  2 \cos^2 (\ell x_1 ) -1 \right) . 
\end{eqnarray}
Furthermore, the expectation value $\langle  T^{00}_{em,2} 
\lbrack  \alpha \rbrack \rangle$ is equal to the energy density of the
free electromagnetic field. Indeed, it holds for the second derivatives
\begin{eqnarray}
   \left.
\delta^2 Z_{em} ' / \delta j_\mu (y) \delta j_\nu (x)
         \right|_{j=0, c_0} 
     &=&
 \left.
 {\rm i} D^{\mu \nu}_1 (x,y)  Z_{em} '  
  \right|_{j=0, c_0} 
\end{eqnarray}
where $ D^{\mu \nu}_1 (x,y)$ tends to the free propagator $D^{\mu \nu}
(x,y)$ in the infinite volume limit, since the last term on the
r.h.s. of Eq. (\ref{modpro}) is of the order $\Omega^{-1}$.
Consequently,  the pure electromagnetic contribution
to the Casimir energy density is given by
\begin{eqnarray}
  e_{em} (x) &=& 
    \langle 
         T^{00}_{em,0} \lbrack {\bar A} \rbrack 
     +  T^{00}_\lambda  \lbrack {\bar A}, c_0 \rbrack 
            \rangle
   \nonumber\\
  &=&
  \frac{1}{2} a^2 \ell^2 \cos^2 (\ell x_1 ) .
\end{eqnarray}

It is more cumbersome to evaluate the fermionic contribution $e_F (x)
= \langle T^{00}_F \rangle - \langle T^{00}_{free} \rangle_{0} $ to the
Casimir energy density, where the energy density of the free Dirac
vacuum is subtracted. Since QED in
dimensions $1+1$ is superrenormalizable, the difference turns out
to be UV finite without further renormalizations.  
We perform the evaluation of the fermionic part of the Casimir energy
in the second quantized formalism. Then we can write
$e_F =  \langle 0 | :  T^{00}_F : | 0 \rangle$ where $ |0 \rangle$ is
the `interacting' vacuum (in the presence of the periodic background field) 
and $: \ldots :$ denotes normal ordering with respect to the normal
vacuum (in the absence of the background field).
The evaluation is performed     
in the following steps:
\begin{enumerate}
\item The fermion field $\psi$ is expanded in terms of the
  eigenspinors $f^{(ks)} (x_1)$ and $g^{(ks)} (x_1)$
 of the Dirac Hamiltonian $H_D ( {\bar A} )$ belonging to the energy
 eigenvalues $\epsilon_{ks}$ and $-\epsilon_{ks}$, resp. Here the
quasi-momentum $k \in \lbrack - \ell/2, \ell/2 \rbrack$ and the integer
$s \ge 0$ enumerating the bands are introduced.

\item Then the creation and annihilation operators $a_{ks}^\dagger$, 
$ b_{ks}^\dagger$ and   $a_{ks}$, $b_{ks}$ of these
 stationary  single particle states are expressed as linear
 combinations of the
  creation and a\-nni\-hila\-tion operators $A_{ks}^\dagger$, $B_{ks}^\dagger$ 
and  $A_{ks}$, $B_{ks}$ of the free 
fermion states $F^{(kr)} (x_1)$ and $G^{(kr)} (x_1)$ of energies
$\epsilon_{kr}^{(0)}  = \sqrt{ m^2 + (k + \ell r)^2}$ and
 $-\epsilon_{kr}^{(0)}$, resp.
  In terms of the latter, the normal ordering is performed.
\item Finally, the normal ordered operator is reexpressed in terms of
  the operators $a_{ks}^\dagger$, $b_{ks}^\dagger$ and $a_{ks}$, $b_{ks}$  
and the vacuum expectation value with
  respect to the vacuum in the presence of the background field is
  taken.
\end{enumerate}
Thus, one arrives to the following expression
\begin{eqnarray}
  T^{00}_F (x)
 &=&   e_F  (x) +
     \sum_{\rho \rho'} \left\{ 
   a_{\rho '}^\dagger (t) a_\rho (t)  {\tilde B}_+^{\rho '}\cdot
   B_-^\rho
 +  b_{\rho }^\dagger (t) b_{\rho'} (t)  {\tilde A}_+^{\rho '}\cdot
   A_-^\rho  
           \right.
         \nonumber\\
    &   &
          \left. 
 +  b_{\rho '} (t) a_{\rho} (t)  {\tilde A}_+^{\rho '}\cdot
   B_-^\rho  
  +  a_{\rho '}^\dagger (t) b_{\rho}^\dagger (t)  {\tilde B}_+^{\rho '}\cdot
   A_-^\rho             
          \right\}
\end{eqnarray}
where, with  $\rho \equiv (ps)$, $\rho' \equiv (p's')$,
\begin{eqnarray}
    {\tilde A}_+^{\rho'} = {\tilde \alpha}_F^{\dagger \; \rho'}
                       +    {\tilde \alpha}_G^{\dagger \; \rho'},
      \qquad
   {\tilde B}_+^{\rho'} = {\tilde \beta}_F^{\dagger \; \rho'}
                       +    {\tilde \beta}_G^{\dagger \; \rho'},      
       \nonumber\\
      { A}_-^{\rho} = { \alpha}_F^{ \rho}
                       -    { \alpha}_G^{\rho},
      \qquad
   { B}_-^{\rho} = { \beta}_F^{ \rho}
                       -    { \beta}_G^{\rho},
\end{eqnarray}
and
\begin{eqnarray}
  \alpha_F^{ps} = \sum_{kr} \epsilon^0_{kr} \alpha_-^{pksr} F_{kr}
   (x),
         \qquad
 \alpha_G^{ps} = \sum_{kr} \epsilon^0_{kr} \alpha_+^{pksr} G_{kr}
  (x),
            \nonumber\\
  \beta_F^{ps} = \sum_{kr} \epsilon^0_{kr} \beta_-^{pksr} F_{kr}
  (x),
         \qquad
 \beta_G^{ps} = \sum_{kr} \epsilon^0_{kr} \beta_+^{pksr} G_{kr}
  (x),
         \nonumber\\
{\tilde \alpha}_F^{* \; ps} = \sum_{kr}  \alpha_-^{* \; pksr} F_{kr}^*
  (x),
         \qquad
{\tilde \alpha}_G^{*\; ps} = \sum_{kr} \alpha_+^{*   \;pksr}  G_{kr}^*
  (x),
            \nonumber\\
 {\tilde \beta}_F^{* \; ps} = \sum_{kr} \beta_-^{* \;pksr} F_{kr}^*
  (x),
         \qquad
 {\tilde \beta}_G^{* \; ps} = \sum_{kr} \beta_+^{* \; pksr} G_{kr}^*
  (x),
\end{eqnarray}
with the constant coefficients
\begin{eqnarray}
  \alpha_-^{pksr} = \int dx F^{* \; kr} (x) g^{ps} (x) ,
        \qquad
  \alpha_+^{pksr} = \int dx G^{* \; kr} (x) g^{ps} (x) ,
         \nonumber\\
  \beta_-^{pksr} = \int dx F^{* \; kr} (x) f^{ps} (x) ,
        \qquad
  \beta_+^{pksr} = \int dx G^{* \; kr} (x) f^{ps} (x) .
\end{eqnarray}
The latter are the overlap integrals of the eigenspinors in the
presence of the background and those in the absence of the background.
Furthermore, the time dependent creation-annihilation operators
 are introduced:
\begin{eqnarray}
  a_{ps} (t) \equiv a_{ps} e^{ - {\rm i} \epsilon_{ps} t } ,
           \qquad
   b_{ps} (t) \equiv b_{ps} e^{ - {\rm i} \epsilon_{ps} t } .
\end{eqnarray}
The constant term 
\begin{eqnarray}
\label{eF}
  e_F (x) = 
  \sum_{ps} \left(
    {\tilde \beta}_G^{* \; ps} \cdot \beta_G^{ps}
  +  {\tilde \alpha}_F^{* \; ps} \cdot \alpha_F^{ps}
    +  {\tilde \alpha}_G^{* \; ps} \cdot \alpha_F^{ps}
    -  {\tilde \alpha}_F^{* \; ps} \cdot \alpha_G^{ps}
           \right) 
\end{eqnarray}
represents the Casimir energy density of the fermion vacuum.
The last two terms of Eq. (\ref{eF}) cancel. 

Performing the calculation results in:
\begin{eqnarray}
\label{enden}
e_F(x)&=&\sum_{
              k\in [-\frac{l}2,\frac{l}2] 
              rr^\prime\in N
             }
\biggl[\epsilon^0_{-kr}
({\cal F}_1^{(krr^\prime)}+{\cal F}_2^{(krr^\prime)})
\cos((r-r^\prime)\ell x) \nonumber\\
&&+(\epsilon^0_{k-r^\prime}-\epsilon^0_{kr})
{\cal F}_3^{(krr^\prime)}
\cos((r+r^\prime)\ell x) \biggr],
\end{eqnarray}
where we introduced the following notations:
\begin{eqnarray}
{\cal F}_1^{(krr^\prime)}&=&\sum_{s\in Z} 
U^{(-kr^\prime)}_\alpha U^{(-kr)}_\alpha 
U^{(-kr^\prime)}_\beta v^{(-r^\prime k s)}_\beta
U^{(-kr)}_\gamma v^{(-r k s)}_\gamma\nonumber\\
{\cal F}_2^{(krr^\prime)}&=&\sum_{s\in Z}
 V^{(-kr)}_\alpha V^{(-kr^\prime)}_\alpha 
V^{(-kr)}_\beta u^{(-r k s)}_\beta
V^{(-kr^\prime)}_\gamma u^{(-r^\prime k s)}_\gamma\nonumber\\
{\cal F}_3^{(krr^\prime)}&=&\sum_{s\in Z}
 U^{(-kr^\prime)}_\alpha V^{(kr)}_\alpha 
U^{(-kr^\prime)}_\beta u^{(-r^\prime k s)}_\beta
V^{(kr)}_\gamma v^{(r k s)}_\gamma.
\end{eqnarray}
$U^{(kr)}$ and $V^{(kr)}$ denotes the eigenspinors of the free
Dirac-equation for the positive and negative energy eigenvalues,
respectively.
\begin{figure}[htb]
\hspace{1cm}
\psfig{file=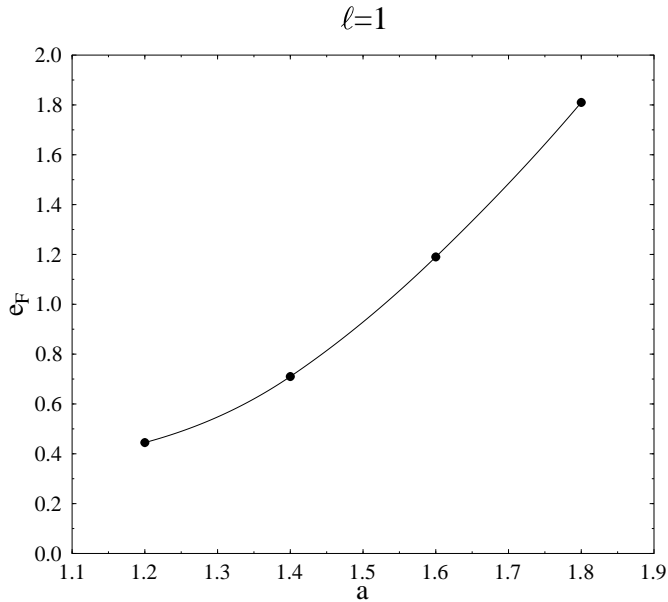,width=10cm}
\vspace{-1cm}
\caption[]{Casimir energy per unit volume of the periodic vacuum in
  $QED_2$.}
\label{thens}
\end{figure}
It is straightforward to establish from Eq. (\ref{enden}) that the volume 
integral of the energy density is not negative:
\begin{equation}
E_F = \sum_{k r s}\biggl\{\epsilon^0_{-k r}\Bigl[
|U^{(-k r)}_\beta v^{(r k s)}_\beta|^2+
|V^{(-k r)}_\beta v^{(r k s)}_\beta|^2\Bigr]\biggr\}\geq 0.
\end{equation}

By numerical calculations we were convinced (see Fig. \ref{thens}) that 
this expression of
the energy only vanishes in case of $a{=}\ell{=}0$.
This means that the Casimir energy $E_c$ of the vacuum is always positive 
if a non-vanishing periodic electric background field is assumed. Thus,
the vacuum of $QED_2$ does not favour a periodic mean field energetically 
with respect to the normal vacuum.

\section{Necessary condition of energy minimum}
The vacuum expectation value ${\bar A}^0 (x)$ is defined by the
minimum of the energy functional 
\begin{eqnarray}
   T E \lbrack {\bar A}^0 (x) \rbrack 
         &=&
   \int dx \langle T^{00} (x) \rangle 
       =
   \int dx \left\langle \left( T^{00}_{em} (x) + T^{00}_{\lambda} (x)
                   + T^{00}_F (x) \right)
             \right\rangle .
\end{eqnarray}
The second term vanishes due to the explicit value of $c_0$. Thus, the
necessary condition of the energy minimum takes the form
\begin{eqnarray}
   \frac{\delta}{\delta {\bar A}^0 (x) } 
         \int dy  \left(  T^{00}_{em,0} \lbrack {\bar A}^0 (y) \rbrack 
                   +
             \left\langle  T^{00}_F (y)  \right\rangle
                    \right)=0 .
\end{eqnarray}
The functional derivative of the first term gives
\begin{eqnarray}
   \frac{\delta}{\delta {\bar A}^0 (x) }   
 \int dy  \left(  T^{00}_{em,0} \lbrack {\bar A}^0 (y) \rbrack \right)
       &=&
   - \nabla^2 {\bar A}^0 (x) .
\end{eqnarray}
At  one-loop order the fermionic contribution to the energy can be 
rewritten as
\begin{eqnarray}
  TE_F &=& 
 \int dy  \left\langle  T^{00}_F (y)  \right\rangle
        \nonumber\\
     &=&
   Z_F^{-1} \lbrack {\bar A},0, 0 \rbrack \int dy  T^{00}_{F,op} (y)
    \left.    Z_F \lbrack {\bar A}, \eta , {\bar \eta} \rbrack  
    \right|_{\eta={\bar \eta}=0} .
\end{eqnarray}
Taking its functional derivative we find
\begin{eqnarray}
\label{funcder}
  \frac{\delta}{ \delta {\bar A}^0 (x) } TE_F
        &=&
   -  \frac{\delta \ln Z_F \lbrack {\bar A}, 0,0\rbrack}{
          \delta {\bar A}^0 (x) } TE_F \nonumber\\
    &&
     + Z_F^{-1}  \lbrack {\bar A},0, 0 \rbrack \int dy  T^{00}_{F,op} (y)
  \left.  \frac{\delta   Z_F \lbrack {\bar A}, \eta , {\bar \eta}
      \rbrack  
          }{ \delta {\bar A}^0 (x) }
 \right|_{\eta={\bar \eta}=0}
           \nonumber\\
    &=&
  {\rm i} \int dy \biggl[ 
    \frac{\rm i}{2} \partial_y^0 \biggl( 
          \frac{\delta }{ \delta \eta (y) } \gamma^0 \frac{\delta
            }{\delta {\bar \eta} (y) } \biggr)\nonumber\\
   &&
   - \frac{\delta }{\delta \eta (y) } \gamma^0 H_{D0}^y 
     \frac{\delta }{ \delta {\bar \eta} (y) }
                  \biggr]
         \left( {\bar \eta} \frac{ \delta G}{ 
                 \delta {\bar A}^0 (x) }
                 \eta  \right)
          e^{ {\rm i} ( {\bar \eta} G \eta )} \biggl|_0
              \nonumber\\
      &=&
  - {\rm i} {\mbox { tr }} \left( \gamma^0 G(x,x) \right)
     + c^0 (x)  
\end{eqnarray}
with
\begin{eqnarray}
  c_0 (x) &=&  \frac{1}{2} \int dy [
      {\mbox { tr }} ( \gamma^0 G(y,x) \gamma^0 D^0_y
        G(x,y)  )\nonumber\\
&&   
 -   {\mbox { tr }} ( \gamma^0 G(x,y) \gamma^0 D^{0*}_y
        G(y,x) )] .
\end{eqnarray}
Thus, we find the following equation for the field ${\bar A}^0 (x)$
minimizing the energy:
\begin{eqnarray}
   \partial_\nu {\bar F}^{0\nu} 
         &=&
    \left\langle {\bar \psi} (x) \gamma^0 \psi (x) \right\rangle 
    +  c^0 (x) .
\label{mfeq}
\end{eqnarray}
The first term on r.h.s. is the expectation value of the charge
density $j^0=-i{\mbox{tr}}(\gamma^0G(x,x))$, determined via the
propagator $G(x,x)$ as a given functional of $\bar A^0(x)$, and a similar
statement holds for $c^0(x)$.

A similar equation appears in the classical case when a certain 
charge distribution moves in an electromagnetic field. Integrating out
the effect of the charged particle distribution it will result in a 
polarisation charge density term beside the common charge
density in the equation of motion for the scalar potential.
In Eq. (\ref{mfeq}) the term $c^0(x)$ corresponds to a
polarisation charge density, so minimizing the energy with respect to $\bar
A^0(x)$ leads to an equation similar to that of Poisson's
equation in a polarised medium.
Thus Eq. (\ref{mfeq}) can be solved directly for $\bar A^0(x)$ in
principle without the need for solving any other equations. On the
contrary to this
the mean field equation (\ref{mfeqact}) obtained by extremizing the
effective action does not take the polarisation of the vacuum into
account, in order to do this we have to solve a system of operator
equations as well.

\section{Conclusions}

The energy density of $QED_2$ in the presence of a periodic mean field 
is determined from the energy-momentum tensor. For this purpose a  
projection method is worked out which is applied to treat the mean 
electromagnetic field self-consistently. The projected effective action 
and the energy density of the vacuum are derived at one-loop order,
whereas the interaction of the electron-positron field with the
periodic mean field is treated exactly. It is established,
that the negative of the effective action must not be regarded as the energy 
of the system if the background field is not constant.
It was shown that the necessary conditions of the extremum of the 
effective action and the minimum of the energy lead to different
equations for the vacuum expectation value of $\bar A^0$.
Eq. (\ref{mfeqact}) obtained by extremizing the effective action is
the vacuum expectation value of the Poisson's equation that does not
include the polarisation of the vacuum due to one-loop 
radiation corrections.
Those are accounted for by separate operator equations for the 
quantum fluctuations of the electromagnetic field and for the Dirac
field.
On the other hand Eq. (\ref{mfeq}) obtained by minimizing the energy
functional includes the polarisation effects of the vacuum.

The expectation value of the component $T^{00}$ of the energy-momentum 
tensor was determined as the function of the amplitude $a$ and wave 
number $l$ of the static periodic scalar potential ${\bar A}^0 (x_1) 
= a \cos(l x_1)$. It is found that the vacuum configuration with this
periodic electric mean field is not favoured energetically compared 
to the normal vacuum. The volume integral of the energy density plotted 
against the amplitude $a$ in Fig. \ref{thens} shows that the energy of 
the system increases with increasing $a$ monotonically.

The result obtained contradicts to our naive expectation discussed in the 
Introduction. Possibly, the reason is that 
the naive picture mentioned there does not take
into account the uncertainty principle i. e. for a certain $\ell$ the
energy spread of a wave packet localized in an interval of $\sim
1/\ell$ could be much larger than the amplitude of the potential
and of course in this case our naive picture is not valid any more.
We set now forth our work looking for periodic ground states at finite
chemical potential.

\appendix

\section{Solution of the Dirac equation in periodic external field}
\label{soldirac}
\def\thesubsection{\thesection.\arabic{subsection}.}
\subsection{Relativistic Bloch waves}

To get the projected effective action at one-loop order in Eq. 
(\ref{effac}) we must find the fermionic single-particle energies,
so we have to solve the Dirac equation in the presence of the potential
(\ref{ansatz}):
\begin{equation}
(i\gamma^{\mu}(\partial_{\mu}-ie\bar A_{\mu})-m)\psi=0.
\label{diracegy}
\end{equation}
We look for the solution of Eq. (\ref{diracegy}) in the form of Bloch waves 
corresponding to the energy eigenvalues $E$:
\begin{eqnarray}
\label{megoldas}
\psi_{\alpha}=e^{-i\epsilon_{ps}t}
e^{ipx}\sum_{n=-\infty}^\infty u_{\alpha n}e^{inlx}
=\left\{\begin{array}{r@{,\quad \mbox{if}\quad}l}
e^{-i\epsilon_{ps}t}f^{ps}(x)& E=\epsilon_{ps}>0 \\
           e^{i\epsilon_{ps}t}g^{ps}(x)& E=-\epsilon_{ps}>0 
           \end{array}\right.
\end{eqnarray}
Inserting (\ref{megoldas}) into (\ref{diracegy}) we find
\begin{equation}
\sum_{n=-\infty}^\infty\biggl(
u_{\alpha n}(\epsilon\gamma^0-(p-nl)\gamma^1-m{\bf 1}) \ + \ \frac{a}2 \gamma^0
(u_{\alpha n+1}+u_{\alpha n-1})\biggr)e^{i(p+nl)x-iEt}=0.
\label{matrix}
\end{equation}
We get non-trivial solutions when the determinant of the matrix 
appearing next to the Dirac-spinors equals zero.
If we had solved the Schr\"odinger equation in the presence of
this sinusoidal potential, the form of the matrix would have been
tridiagonal which means that the diagonal and the neighouring diagonal
elements are nonzero.
In relativistic case the matrix elements are replaced by $d$
dimensional $\gamma$ matrices but the structure of the matrix remains
unchanged.

\subsection{Eigenvalues, eigenspinors}\label{eigenval}

We cannot get the eigenvalues analitically because of the
complicated structure of the matrix but we can determine them as
precisely as we wish.
In numerical calculations we work with matrices with finite dimension. 
Using the well-known identity \cite{numrec}
\begin{equation}
{\rm det}[A]={\rm det}\pmatrix{
P & Q \cr
R & S \cr
}={\rm det}[P] \ {\rm det}[S-RP^{-1}Q]\label{lemma}
\end{equation}
we can reduce the problem of calculating the determinants of these 
$d(2n+1)\times d(2n+1)$ dimensional matrices to calculating four 
dimensional matrices by identifying the upper left matrix element 
of the matrix under investigation with $P$ in Eq. (\ref{lemma}). 
($n$ denotes the number of $u_{\alpha,i}$-s taken into account in 
Eq. (\ref{matrix}).) Using identity (\ref{lemma}) $2n$ times we get a 
product of $2n$ determinants of four dimensional matrices.
We computed the determinant of Eq.(\ref{matrix}) as the function of $E$
and determined its zeros, corresponding to the energy eigenvalues 
$E=\pm\epsilon$

In order to evaluate $-\Gamma$ we have to sum the eigenvalues 
$-\epsilon_{ps}<0$ and extract from it the sum of the negative eigenvalues 
of the free Dirac equation. This difference will depend on the accuracy 
of the eigenvalues, the number of the eigenvalues taken into account
in the sum, the size of the chosen matrix and, of course, the parameters 
of the potential, the amplitude $a$ and the wave number $\ell$. We have to 
be convinced of the stability of the numerical calculation. We increased 
the number of members in the sum and the numerical accuracy of the 
determination of eigenspinors by choosing larger matrices as far as we 
have seen that the energy difference does not change significantly.

To get the expectation value of the component $T^{00}$ of the
energy-momentum tensor we also have to determine the eigenspinors of the
Dirac equation (\ref{diracegy}). We calculated them with the help of the 
eigenvalues by solving a system of homogeneous linear equations 
(\ref{matrix}) for the eigenspinors.

\vskip 10pt
\noindent \large {\bf Acknowledgement}\normalsize
\vskip 10pt
\noindent
The authors would like to thank  J. Pol\'onyi for consulting this work and  
G. Plunien for the valuable discussions. S.N. thanks G. Soff for his
kind hospitality. K.S. expresses 
his gratitude for the follow-up grant of the Alexander von Humboldt Foundation 
and  W. Greiner for his kind hospitality. This work was supported by the 
projects OTKA T023844/97, DAAD-M\"OB 27/1999 and NATO Grant 
PST.CLG.975722.

\vfill\eject

\end{document}